%Paper: gr-qc/9402038
%From: gottlieb@math.purdue.edu (Daniel H. Gottlieb)
%Date: Mon, 21 Feb 94 16:27:41 EST

%amstex, 8 pages
\magnification=\magstep1
\input amstex
\documentstyle{amsppt}
\topmatter
%\title
%A Gravitational Lens need not Produce an Odd Number of Images
%\endtitle
%\author
%Daniel Henry Gottlieb
%\endauthor
%\endtopmatter
%\subhead {1.  Introduction}\endsubhead
\document
\null
\vskip 2truein
\centerline{\bf A GRAVITATIONAL LENS NEED NOT }
\centerline{\bf PRODUCE AN ODD NUMBER OF IMAGES}
\medskip
\centerline{by}
\bigskip
\centerline{\bf Daniel H.\ Gottlieb}
\bigskip
\centerline{\bf Department of Mathematics}
\bigskip
\centerline{\bf Purdue University}
\centerline{\bf  West Lafayette, Indiana}
\vskip .25truein
\noindent
{\bf Abstract}

Given any space-time $M$ without singularities and any event $O$, there is a
natural continuous mapping $f$ of a two dimensional sphere into any space-like
slice $T$ not containing $O$.  The set of future null geodesics (or the set of
past null geodesics) forms a 2-sphere $S^2$ and the map $f$ sends a point in
$S^2$ to the point in $T$ which is the intersection of the corresponding
geodesic with $T$. To require that $f$, which maps a two dimensional space into
a three dimensional space, satisfy the condition that any point in the image of
$f$ has an odd number of preimages, is to place a very strong condition on $f$.
This is exactly what happens in any case where the odd image theorem holds for
a
transparent gravitational lens. It is argued here that this condition on $f$ is
too restrictive to occur in general; and if it appears to hold in a specific
example, then some  $f$ should be calculated either analytically or numerically
to provide either an illustrative example or counterexample.
\vfill\eject
\baselineskip=18pt
\noindent
{\bf Introduction}

Since 1979 astronomers have been looking for an odd number of images in
gravitational lensing events.  There have been many discoveries since the
first event in 1979.  In most cases
{\it only} an even number of discrete images
have been found.  We assert that the topology and geometry of space-time seem
to make it very unlikely that only odd numbers of images exist.  Some of
the topological arguments for an odd number of images are very persuasive,
even though they are based on a Euclidean space-time.  Hence the prevalence
of even number images should be taken as another vindication of Lorentzian
space-time as a model of the universe.

In 1980 C.~C.~Dyer and R.~C.~Roeder \cite{D--R} predicted an odd number of
images for a spherical symmetric transparent lens (i.~e. Galaxy). In 1981
W.~L.~Burke \cite{B}
claimed that there must be an odd number of images for any bounded
transparent lens subject to an assumption that the bending of light rays
decreases as the light rays are far from the lens.  The argument constructed
a vector field on the plane of the lens and showed the index had to be one.
So then, assuming the local index of each zero was $\pm 1$, the number of
zeros had to be odd to add up to the global index of $1$.  Each zero
corresponds
to light rays.

In 1985 R.~H.~McKenzie \cite{Mc} wrote down an argument using the degree of a
map between two $2$ dimensional spheres which asserted that there were an odd
number of images.  This argument needed no assumptions on the amount of
bending and obviously improved Burke's approach.  This argument was widely
known among astronomers and is very convincing.  However it is done in
$3$-space
and not in $4$-dimensional space-time.  McKenzie notes this and then provides
an argument using Morse Theory on $4$-dimensional space-time, applying
correctly Karen Uhlenbeck's version of Morse Theory for Lorentzian Manifolds,
\cite{U}.  It is widely believed today that the necessity of an odd number
of images has been precisely established and that the contradictory evidence
is a result of difficulties of finding the third image,
\cite{P}, although on page 176 of
\cite{S--E--F} they state that McKenzie's conditions are physically obscure.

In this paper we translate the degree argument directly into $4$-dimensional
space-time and we see that an extremely restrictive condition must be true
of the space-time in order to obtain the odd images conclusion.  The condition
is that the pencil of past
null-geodesics from the observer must intersect every
past space-like slice in a $2$-sphere.

We give two examples of $4$-dimensional Lorentzian Manifolds for which this
condition is false.  The first one because of the topology and the second one
because of the geometry.  Then we argue that the conditions under which
McKenzie's Morse Theory argument would apply are even more extremely
restrictive.
\medskip

\subhead{2.  Global lensing in Lorentzian space-time}\endsubhead

We reproduce the topological argument given by McKenzie
on page 1592 of \cite{Mc} which
establishes the odd image result for Euclidean space.  Then we try to
reproduce the argument in Lorentzian space-time.

``There is a relatively simple demonstration of why there are an odd number of
images.  Although it seems to be well known among astronomers it does not
appear
to have been published before and so is given here.  Consider the situation
shown in Fig.~1.  A light source is located at $S$ and an observer at $O$.
There is a transparent galaxy $G$ somewhere between $S$ and $O$.
A map $f$ from the small sphere $A$ to the sphere $B$ is defined as follows.
The map $f$ maps a point $x$ on $A$ to the point on $B$ where the light ray
through $O$ and $x$ intersects $B$.  The number of images of $S$ seen by $O$
is the number of points on $A$ mapped onto $S$.

\vskip2truein
\baselineskip=8pt

\noindent
FIG. 1. {\bf A galaxy $G$ is located somewhere between a light source $S$ and
an
observer $O$.  Because of the gravitational field of the galaxy there may be
more than one light ray from $S$ to $O$.  $f$ maps the sphere $A$ onto the
sphere $B$.  If $x$ is on $A$ then $f(x)$ is defined to be the point on $B$
where the ray through $O$ and $x$ intersects $B$}.

\baselineskip=18pt
\medskip

Suppose $g:M\to N$ is a smooth map between manifolds of the same dimension
and that $M$ is compact.  If $y$ is a regular value of $g$ then we define

$$
\text{deg}(g,y)=\underset{x\in g^{-1}(y)}\to\sum\ sgn\ dg_x,
$$
where $sgn\ dg_x=+1(-1)$ if $dg_x:T_x(M)\to T_y(N)$ preserves (reverses)
orientations.  It turns out that $\text{deg}(g,y)$
is the same for all regular $y$;
it is called the degree of $g$ and denoted  $\text{deg}(g)$.

In an actual physical situation it is reasonable to assume that there will be
a point $y$ on $B$ such that $f^{-1}(y)$ is a single point, i.~e.~ there is
only one ray from $O$ to $y$.  Thus, $\text{deg}(f) = 1$.

Let $n_+(n_-)$ be the number of points $x$ in $f^{-1}(S)$ such that
$sgn\ df_x=+1(-1)$.  Thus, $n_+(n_-)$ is the number  of images of $S$,
seen by $O$, which have the same (opposite) orientation as the source, and
$$
n_+ - n_-=\text{deg}(f,S)=\text{deg}(f)=1.
$$
Thus, if $O$ sees $n=
n_+- n_-$ images of $S$ then $n=2n$, and so $n$ is odd, and the demonstration
is complete.''

Now we consder this argument in Lorentzian space-time, $M$.  We consider every
past directed null geodesics eminating from the observer $O$.  There is one
geodesic for each point in the celestial sphere $A$.  (More precisely, take a
``unit'' sphere in the past null cone of $O$ in the tangent space to $O$.  Then
there is a unique past directed null line for each point of the sphere and the
exponential map maps this line onto a past directed geodesic.)  Now let $T$ be
a space-like slice containing the source $S$.  Then it is natural to assume
that each geodesic intersects $T$ exactly once.  (If this assumption does not
hold it makes the odd image ``theorem'' even more dubious.)  So we can define
a map $f:S^2\to T$.  Now what corresponds to the sphere $B$?  It must be the
image $f(S^2)$ of $S^2$ in the three dimensional manifold $T$.  It seems
unlikely that $f(S^2)$ would be a sphere if $f$ is not injective.  Only if
$f(S^2)$ were a toplogical
sphere would we be entitled to use the degree argument,
otherwise it is invalid.

This would strike any differential topologist or geometer as obvious.  It may
be possible to construct such an $M$, but these $M$'s would be quite special.
\medskip

\subhead{3.  Two Examples}\endsubhead

We give two examples of space-times which do not have the property that pencils
of null geodesics intersect space-like slices in $2$-spheres.  Many more
examples can be constructed using Barrett O'Neill's book \cite{ON},
Corollary 57 on page 89 and warped products on pages 207--209.

\noindent a)
Let $M=S^1\times S^1\times S^1\times{\Bbb R}$.  The universal covering space is
$\widetilde M=\Bbb R^4$.  Let $\widetilde M$ be Minkowski space, so it has the
Minkowski metric.  It induces the same metric on $M$.  The geodesics of
$\widetilde M$ are straight lines and their images are the geodesics of $M$.
Pencils of null geodesics do not intersect space-like slices
in spheres in this $M$.

\noindent b)
Let $M=\Bbb R^4=\Bbb R^2\times\Bbb R^2$.  Let the second $\Bbb R^2$ have the
Minkowski metric.  We will put a Riemannian metric on first $\Bbb R^2$ and then
we take the product metric.  We note that a geodesic of the first $\Bbb R^2$
factor coupled with a time-like line in the second factor is a null geodesic in
$M$, (i.~e.~ if $\alpha:\Bbb R\to\Bbb R^2$ is a geodesic of the first
factor and $\beta:\Bbb R\to\Bbb R^2$ is a
time-like geodesic of the second factor with the same speed as $\alpha$,
then $\alpha\times\beta:\Bbb R\to\Bbb R^2\times\Bbb R^2$ is a null geodesic
of $M$).  So if we produce an $\Bbb R^2$ so that the exponential map of
geodesics eminating from a point $x$ carries some circle in the tangent plane
at $x$ into a set in $\Bbb R^2$ which is not a circle, then the pencil of null
geodesics intersecting a space-like slice in $M$ is not a two sphere.

One can visualize a metric on $\Bbb R^2$ by embedding it as a surface $T$ in
Euclidean $3$-space.  The geodesics are characterized as these paths in $T$
whose acceleration is orthogonal to the surface $T$.  Now it is easy to
construct examples with the desired property.

One that works is the following.  Take an arc of a circle whose length exceeds
a half circle.  Extend the ends of this arc by the tangent lines at the ends
of the arc.
The lines intersect in a Point $A$.  Now take a small interval perpendicular
to the plane in which the curve just constructed, $\gamma$, lies.  Move
this interval along $\gamma$ so that it is perpendicular to the plane over
the arc and so that it lies in the plane along most of the two extended lines
including their intersection $A$.  The interval should be twisted in moving
from the ends of the arc so that the interval sweeps out a smooth surface with
two boundary components.  Then extend this ``old fashioned men's collar''
to a surface $T$ in $\Bbb R^3$.

Let $O$ be the midpoint of the circular arc on $T$.  Then the geodesics
of fixed length greater than $OA$ on $T$ near $\gamma$ clearly do not end in a
circle.

\bigskip
\noindent
FIG. 2
\vskip1.5truein

We can adjust this example so that the nonflat part of $T\times{\Bbb R}$ is
bounded in any space-line slice $T\times {\Bbb R}\times s\subset{\Bbb
R}^2\times
{\Bbb R}\times{\Bbb R}=M$.  The technique for the adjustment is the warped
product construction, which can be found in \cite{ON}.

\medskip
\subhead{4.  Morse Theory}\endsubhead

R.~H.~McKensie in \cite{Mc} ``proves'' the odd image result by applying
Uhlenbeck's version of Morse theory of Lorentzian manifolds \cite{U}. The
relevant theorems are Theorems 4 (which he calls the local theorem) and
Theorem 5 (the global theorem).  The global theorem is less relevant to the
study of gravitational lensing then the local theorem according to
McKensie.  This is the case both for practical considerations of how
observations are made, and because the hypotheses of the global theorem do not
hold in realistic space-time models.

The statement of the local theorem is difficult to understand since
McKensie does not make clear how the points $q$ and $r$ and set $B$ are
related to the history of the source $T$ and the observer $p$.  The most
reasonable interpretation is that $\Omega(T,p)^c$ is a deformation retract of
$\Omega(T,p)$ which he assumes is contractible.  This is a wordy way of
assuming that $\Omega(T,p)^c$ is contractible.  Now as $c$ varies,
$\Omega(T,p)^c$
will not be contractible in general since every time $c$ passes through a
critical value of $T$, the topology of $\Omega(T,p)^c$ is altered by
attaching a cell (which corresponds to a new geodesic from $T$ to $p$).
But it is impossible to attach only one cell to a contractible space and
still have it be contractible.  Thus for ``most'' $c$ the hypothesis is not
true unless there are pairs of geodesics from $T$ to $p$
for each critical value for $c$.

\medskip
\subhead{5.  Discussion}\endsubhead

The fact that there is no odd image theorem shows that even
for very mild lensing the Euclidean approximation is fundamentally wrong. As
a general rule, every Euclidean argument should be looked at in Lorentzian
space-time to see if it can be reproduced there in principle. If the argument
cannot be reproduced, then its consequences are a test for the Lorentzian
model of the universe.

The argument does not really depend on mapping the entire two sphere
into a spacelike slice. If a portion of the light cone two-sphere is
mapped into  a space-like slice, it would give rise, with the most usual
choices, to a mapping $f$ of a plane into three space. The requirement that
the mapping have only an odd number of points in each coincidence is
still a strong condition. For convincing special cases where it appears
the odd images theorem should hold, the mapping  $f$ should be
calculated, either analytically or by ray tracing with a computer. Then
if the odd image property is true for that $f$, we may be able to infer
properties for such mappings in odd imaging cases.
\vfill\eject

\end